\documentclass[aps,prb,preprint]{revtex4-2}

\usepackage{amsmath,amssymb}
\usepackage{graphicx}
\usepackage{epsfig}
\usepackage{multirow}
\usepackage{color}

\newcommand{\vect}[1]{\boldsymbol{#1}}
\newcommand{\bhat}[1]{\widehat{\boldsymbol{#1}}}

\newcommand{\zhat}{\bhat{z}}
\newcommand{\rhohat}{\bhat{\rho}}
\newcommand{\phihat}{\bhat{\varphi}}

\begin{document}

\input{epsf}

\title{A Modern Interpretation of Newton's Theorem of Revolving Orbits}
\author{Nolan Samboy and Joseph Gallant}

\affiliation{Department of Physical and Biological Sciences,\\ Western New England University,\\
            1215 Wilbraham Road, Springfield, MA 01119}

\date{\today}
\begin{abstract}
Newton's \textit{Theorem of Revolving Orbits} derives the force that is necessary to explain
a particular precession that leaves the shape of an orbit unchanged.
Newton showed that for an orbiting body that is already subject to any central force, 
the inclusion of an additional inverse-cube central force will change that body's angular speed without affecting its 
radial motion; this leads to the orbit's precession. After reviewing
the relevant concepts of Kepler orbits, we present a full description and explanation of Newton's theorem using modern physical 
and mathematical approaches aimed at a general audience. We specifically highlight the use of the transformation between
a rotating reference frame and an inertial reference frame, making this a suitable topic for an upper-level 
undergraduate mechanics course. 
\end{abstract}

\maketitle
\section{Introduction}
\label{sec:intro}
In his \textit{Theorem of Revolving Orbits}, Newton considered the problem of orbiting
celestial bodies that revolve (or precess) about what Newton called the `force center' of the orbit.~\cite{Principia}
At this point in the \textit{Principia}, Newton was focused on merely deriving the mathematical 
form of the force that was needed to describe precession; he was not concerned with the physical origin of that force. 
Thus, he assumed that the central force acting on the orbiting body was directed towards a single point, 
i.e. the force center. 
Newton determined that any orbit (defined by any central force) would precess if an additional central force,
proportional to the inverse-cube of the orbiting body's radial distance from the force center, acted on the
orbiting body. As
Newton established this theorem simply as a stepping stone to be able to analyze nearly circular orbits and then
the precession of the moon, it is often overlooked and thus (as noted by Chandrasekhar) 
``not generally known.''~\cite{Chandra} 

While the earliest references of this theorem appear in the classic texts of Whittaker and Lamb~\cite{Whittaker,Lamb}, 
Lynden-Bell has most famously explored the various implications,
extending the framework well beyond Newton's original propositions.~\cite{Bell-shapes,Bell-Jin,Bell-Hamilton} 
More recently, there have also been works that incorporate General Relativity 
effects into the problem, further extending the basic theory.~\cite{Nguyen,Christian} As these works tend to
be technical astronomy/cosmology papers, they mostly eschew the elementary details of the original 
problem in favor of more rigorous results. 
However, we believe that the omitted technical details could be of great value to those who are unfamiliar
with the theorem and want to understand how to properly apply the arguments.
Our goal in this paper is to present Newton's \textit{Theorem of Revolving Orbits} using modern techniques, 
while also providing
the context that could help make this topic part of an upper-level undergraduate mechanics course. 
\section{Newton's Theorem of Revolving Orbits}
\label{sec:theorem}
Newton's \textit{Theorem of Revolving Orbits} primarily refers to Propositions 43 and 44 in Section IX of Book I of the 
\textit{Principia}.
By this point, Newton had defined his three laws of motion and he had
established the relationship between an inverse-square central force and the resulting conic-section orbits.
We will first quickly review basic orbital mechanics as needed for application to
the revolving orbit theorem. We again note that
Newton is describing bodies that orbit about a fixed point of infinite mass, assumed to be at the origin of the
coordinate system. He does not discuss the two-body, center-of-mass problem until Section XI.
\subsection{Preliminaries}
\label{subs:prelim}
As the two-body problem is constrained to a 2-D plane, we let $z=0$ and present the problem in cylindrical polar coordinates
($\rho$, $\varphi$), where $\rho$ is the radial distance from the central force point and $\varphi$ is the azimuthal angle. 
In these coordinates, the position, velocity, and acceleration of any orbiting body are respectively given by:~\cite{Symon}
\begin{align}
\label{eq:pos}
\vect{r}&=\rho\rhohat \\
\label{eq:velo}
\vect{v}&= \dot{\rho}\rhohat +\rho\dot{\varphi}\phihat\\
\label{eq:accel}
\vect{a}&= (\ddot{\rho}-\rho\dot{\varphi}^2)\rhohat + (\rho\ddot{\varphi}+2\dot{\rho}\dot{\varphi})\phihat
\end{align}
For any central/radial force $\vect{F}(\rho) = \pm F(\rho)\rhohat$, the $\phihat$ term in Eq.~\eqref{eq:accel} 
must be zero, and thus this $\phihat$ term can be trivially integrated to yield $\rho^2\dot{\varphi}\equiv$ constant. 
This is, of course, a statement
that the angular momentum $\ell = m\rho^2\dot{\varphi}$ of the orbiting body is a constant of the motion. The area
law immediately follows from this condition: Since the triangular area swept out by an orbiting body in a small
time $dt$ is given by $dA = \frac{1}{2}\rho(\rho d\varphi)$, then
\begin{equation}
\label{eq:arealaw}
\frac{dA}{dt} = \frac{\rho^2}{2}\dot{\varphi} = \frac{\ell}{2m}\,\,.
\end{equation}
We again point out that Eq.~\eqref{eq:arealaw} holds for \underline{any} central force, 
not just the inverse-square force. 

From Newton's second law, 
the equation of motion describing the orbiting body's path is given for any central force to be
\begin{equation}
(\ddot{\rho}-\rho\dot{\varphi}^2)\,\rhohat = \frac{\vect{F}(\rho)}{m}\,\,.
\end{equation}
A convenient approach to solving this equation is to recast it such that the radial distance depends on the angle
and not the time.~\cite{Symon,Goldstein} 
The solutions then give the shapes of the orbital trajectories, which are often more desired than the
explicit time-dependence of the coordinates. For central forces that go as $\sim \rho^{-n}$, 
this is most easily accomplished
by letting $u=1/\rho$ and using the chain rule $d/dt = \dot{\varphi}\,d/d\varphi$, where
$\dot{\varphi} = \ell u^2/m$ from above.
We quote the result here, but the derivation is provided in Appendix~\ref{appA}:
\begin{equation}
\label{eq:Binet}
-\frac{\ell^2u^2}{m}\left(\frac{d^2u}{d\varphi^2} + u\right)\rhohat = \vect{F}(1/u)\,\,.
\end{equation}
This equation allows us to either solve for the orbit $u(\varphi) = 1/\rho(\varphi)$ corresponding
to a given force, or solve for the force given a particular orbit.
While much of Newton's work on orbital mechanics applies to any central force and the orbits that force produces, 
he spends a great deal of time discussing elliptical orbits and often uses such 
elliptical orbits as his example cases.
In the specific case of the Kepler problem, $\vect{F}(1/u) = -\beta u^2$~$\rhohat$, 
and Eq.~\eqref{eq:Binet} reduces to
\begin{equation}
\label{eq:Binet-Kep}
\frac{d^2u}{d\varphi^2} + u = \frac{m\beta}{\ell^2}\,\,.
\end{equation}
The orbit is then found to be
\begin{equation}
\label{eq:keporb}
\rho(\varphi) = \frac{\ell^2/(\beta\,m)}{1+\left(\dfrac{\ell^2}{\beta\,m\,\rho_{\rm max}}-1\right)\cos{\varphi}}\,\,,
\end{equation}
where we define $\rho$ to be a maximum at $\varphi= 0$ for convenience.
The term in parenthesis is known as the eccentricity of the orbit $\varepsilon$ and is often alternatively expressed as
$\varepsilon~=~\sqrt{1+2E\ell^2/m\beta^2}$, where $E$ is the total energy of the orbit.
We note that for $0 < \varepsilon < 1$, Eq.~\eqref{eq:keporb} defines an ellipse.

\subsection{Proposition 43}
\label{subs:prop43}
In Proposition 43, Newton states 
\begin{quote}
\textit{It is required to find the force that makes a body 
capable of moving in any trajectory that is revolving [precessing]
about the center of forces in the same way as another body in that same
trajectory at rest.}\cite{Principia}
\end{quote}
In other words, Newton wanted to consider an orbit that precessed around the force center, but was otherwise identical 
to the static orbit; i.e. if the precessing orbit were viewed from a co-rotating reference frame, it
would have the exact shape and size as the static orbit.
\begin{figure}[h]
\includegraphics[width=6in]{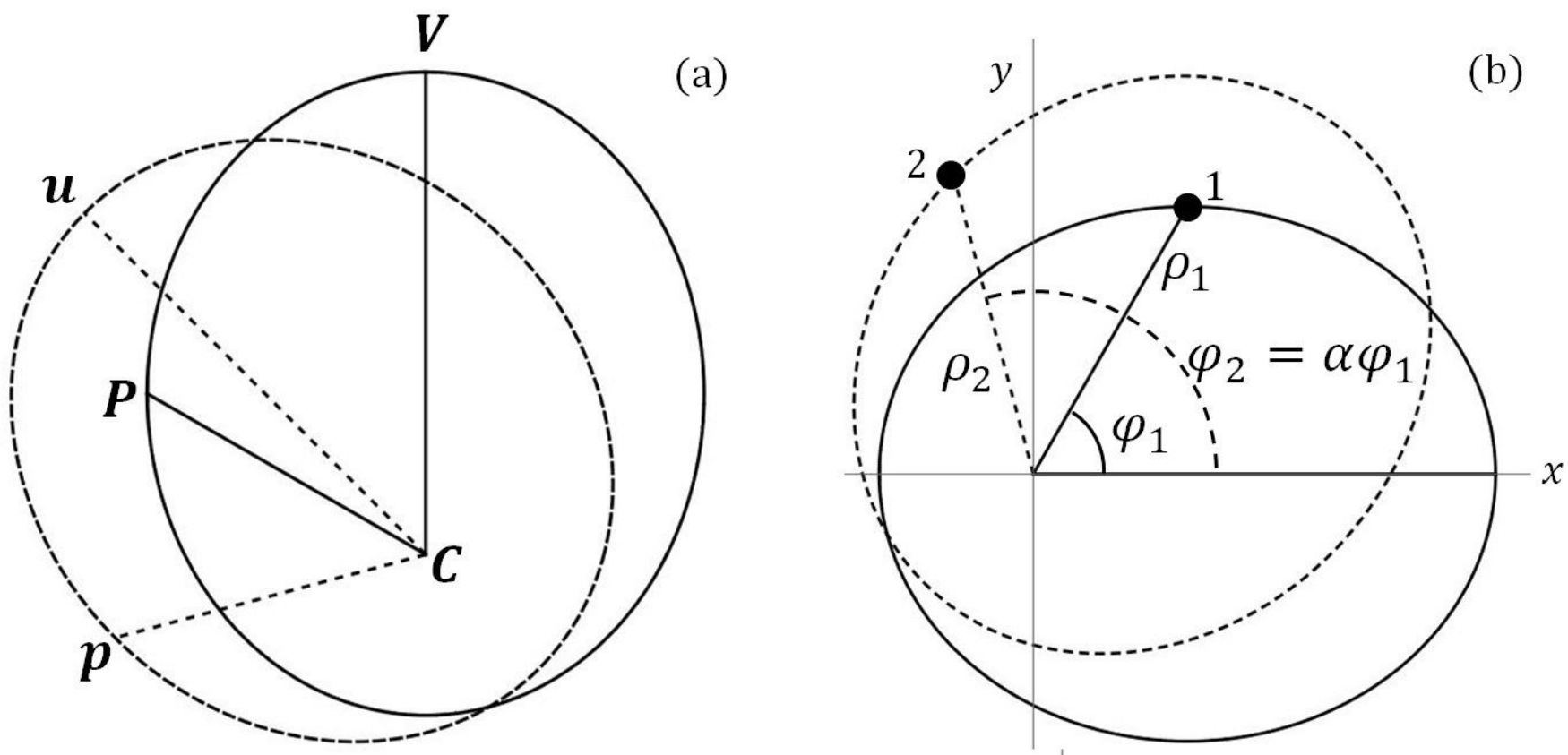}
\caption{Newton's Revolving Orbits: Panel (a) is a reproduction of Newton's figure in the \textit{Principia}, 
         while panel (b) projects
         the figure onto fixed axes, while also defining the polar coordinates for body 1 (follows the solid curve) and
				 body 2 (follows the dashed curve). In both panels,
				 the static orbit is given by the solid curve, and the precessing orbit is given by the dashed curve.
				 While neither panel represents the actual orbit path of the precessing body, they provide a qualitative
				 picture of the scenario that Newton considered: as a body precesses about the force center $C$, the elliptical
				 shape of its orbit does not change.}
\label{fig:NewtonOrbits}
\end{figure}
Fig.~\ref{fig:NewtonOrbits}(a) reproduces Newton's figure from the \textit{Principia}, where
the solid ellipse represents the static orbit, the dashed ellipse represents the revolving orbit, and $C$ is the 
force center for both orbits. Note that the dashed ellipse is the same size and shape as the solid ellipse, it is merely rotated;
thus, the arc $VP$ is equal to the arc $up$, and so we designate $p$ and $P$ as ``corresponding points'' on 
the two ellipses.

Although this depiction is not physically accurate, Newton used the image to visually convey
the scenario that he was looking to define: as the ellipse
revolves [precesses] around point $C$, its shape does not change.
Since the shape of the ellipse does not change, the static orbit and the precessing orbit have the same eccentricities.
The precessing orbit, however, has additional angular momentum coming
exclusively from the precession. 
Summarizing Newton's meticulous geometric proofs: for any two ``corresponding points'' $p$ and $P$,
if segment $Cp$ is equal to the segment $CP$, then
angle $\angle VCp=\alpha\,\angle VCP$ (where $\alpha$ is a constant), and so a body in the static
orbit moves from $V$ to $P$ in the same amount of time that a body in the precessing orbit moves from
$V$ to $p$.

In Fig.~\ref{fig:NewtonOrbits}(b), 
we replicate Newton's diagram using a modern description, 
where $(\rho_1,\varphi_1)$ is the position of body 1 in the static orbit and 
$(\rho_2, \varphi_2)$ is the position of body 2 in the precessing orbit, with both positions measured 
from a fixed reference frame. Newton argued that 
if $\rho_2(\varphi_2)=\rho_1(\varphi_1)$ for any two ``corresponding points,'' then
$\varphi_2 = \alpha\varphi_1$ for all ``corresponding points,'' where $\alpha$ is a constant. 
Thus from the area law~\eqref{eq:arealaw}, we have
\begin{equation}
\label{eq:arealaw2}
\frac{dA_2}{dt}=\frac{\rho_2^2}{2}\dot{\varphi_2}=\frac{\rho_1^2}{2}\alpha\dot{\varphi_1} = \frac{\alpha\ell_1}{2m}\,\,,
\end{equation}
where $\ell_1$ is the angular momentum of the static orbit.
Given that $\alpha$ is constant, the precessing orbit also sweeps out equal areas in equal times, with a constant
\underline{total} angular momentum given by $\ell_2 = \alpha\ell_1$. 
This is a subtle point that occasionally gets overlooked: while $\ell_2=\alpha\ell_1$ defines
the total angular momentum of the precessing orbit, this total value reflects the angular momentum
coming from the elliptical orbit (equal to that of the static orbit) plus the angular momentum coming from
the precession:
\begin{equation}
\label{eq:angmom}
\vect{\ell}_2 = \vect{\ell}_1 + \vect{\ell}_{\rm pre}\,\,.
\end{equation}
Whether this $\ell_2$ is larger or smaller than $\ell_1$ depends on the value of $\alpha$ and whether the precession is in 
the same direction as the orbit or in the opposite direction as the orbit (see below). 
Either way, $\ell_2=\alpha\ell_1$ is a constant of the motion, and so
the force required to produce the precessing orbit must be a central force.
\subsection{Proposition 44}
\label{subs:prop44}
Having argued in Proposition 43 that only a central force was capable of producing a precessing orbit such that  
$\rho_2(\alpha\varphi)=\rho(\varphi)$, Newton next showed in Proposition 44 that the difference between the force producing
the precessing orbit $\vect{F}_2(\rho)$ and the force producing the static orbit $\vect{F}_1(\rho)$
must be proportional to the inverse cube of the radial distance from the force center, i.e.
\begin{equation}
\label{eq:invcube}
\vect{F}_2(\rho) - \vect{F}_1(\rho)\propto \rho^{-3}\,\rhohat\,\,.
\end{equation}
To quote Newton:
\begin{quote}
\textit{The difference between the forces under the action of which two bodies are able to move equally - one in an orbit
that is at rest and the other in an identical orbit that is revolving [precessing] - is inversely as the cube of their 
common height [radial distance].}~\cite{Principia}
\end{quote}

Newton used extensive geometric arguments to arrive at his final result, but
using Eq.~\eqref{eq:Binet} and the chain rule to determine the force 
difference has effectively become the standard derivation.~\cite{Chandra,Whittaker,Lamb} 
We will take a slightly different approach, however, and find this force difference by transforming
into a reference frame that rotates with the precessing orbit. While this may seem unnecessary, we believe that 
there is insight to be gained from this approach, and further, it serves as an interesting application for students
in an upper-level mechanics course. Although Newton did not explicitly discuss a rotating frame of reference, the
scenario he described is indeed the one in which the precessing orbit appears static when viewed from a co-rotating
reference frame.

\begin{figure}[h]
\includegraphics[width=4in]{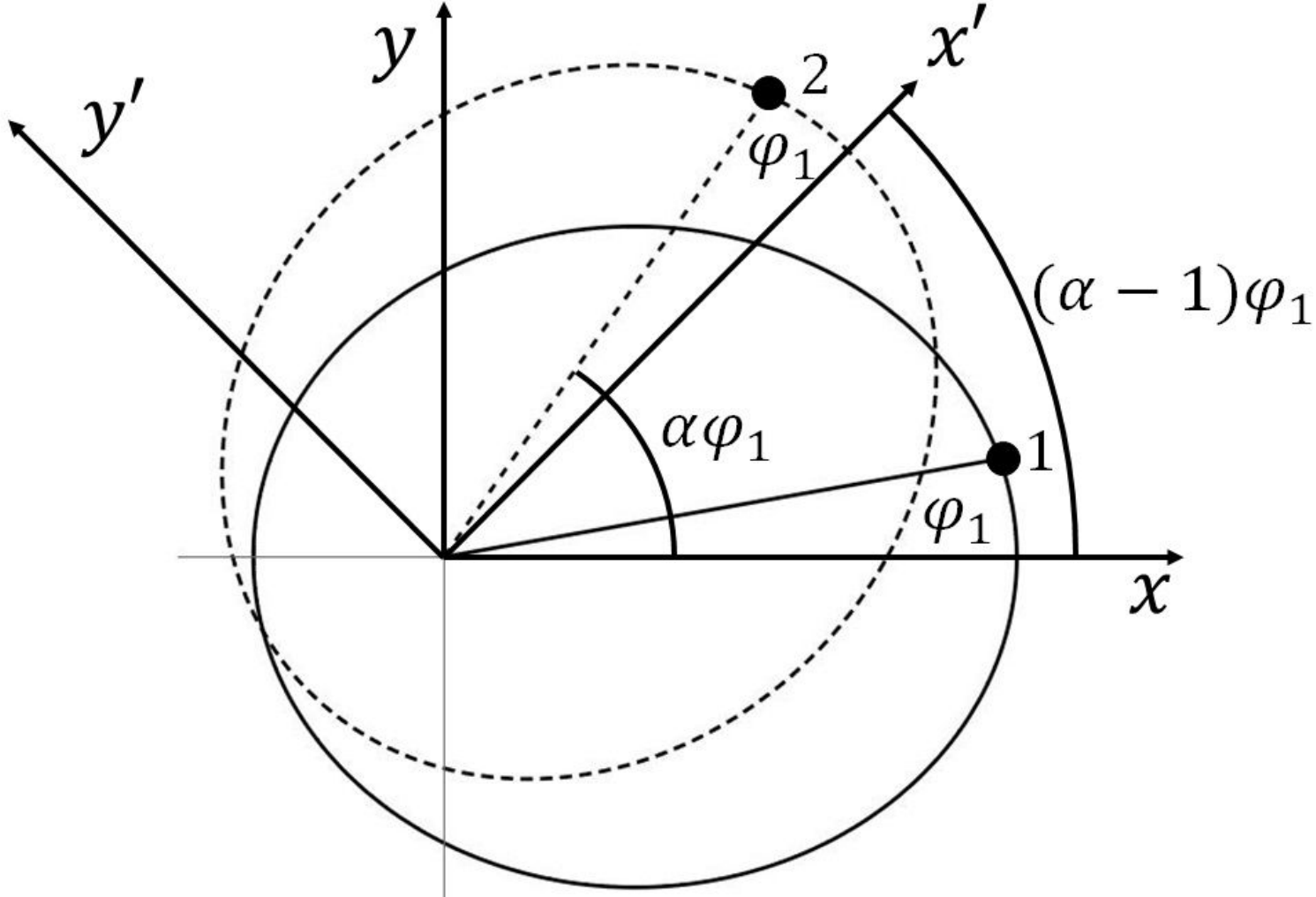}
\caption{Newton's fixed orbit (solid) and revolving orbit (dashed) with two sets of coordinate axes. The $x$, $y$ 
         axes are fixed, while the $x'$, $y'$ axes revolve at the rate $(\alpha-1)\,\dot{\varphi}$ (see text).
				 While body 2 appears to 
				 precess in the fixed frame, it will appear to be a static orbit (of the same size and shape as body 1)
				 when viewed from the revolving frame.}
\label{fig:Revaxes}
\end{figure}

In Fig.~\ref{fig:Revaxes}, we again show the static ellipse and the precessing ellipse, 
but we attach a set of axes $x'$, $y'$ that revolve with the precessing ellipse.
For convenience, we let the $x'$-axis align with the apoapsis (point at which $\rho$ is a maximum) 
of the precessing
ellipse, which makes an angle of $(\alpha-1)\varphi_1$ with the fixed $x$-axis. 
We see that while body 2 makes an angle 
$\alpha\varphi_1$ with the fixed $x$-axis, it makes the angle $\varphi_1$ with the revolving $x'$-axis.
This means that if the $x'$-$y'$ axes rotate at the rate $(\alpha-1)\dot{\varphi_1}$, 
the orbit of body 2 viewed from these primed
axes looks identical to the orbit of body 1 as viewed from the unprimed axes. 
In other words, the effective force acting on body 2 in the revolving
frame $\vect{F}'_{\rm 2}(\rho)$ must be identical to the force acting on body 1 in the static frame 
$\vect{F}_1(\rho)$, i.e. 
\begin{equation}
\label{eq:Forces}
\vect{F}_1(\rho) = \vect{F}'_{\rm 2}(\rho).
\end{equation}
The force on body 2 in the static frame $\vect{F}_2$ can then be found from the standard transformation:~\cite{Symon}
\begin{equation}
\label{eq:transform}
\vect{F}_2(\rho) = \vect{F}'_2 (\rho)+ 
m\,\dot{\vect{\omega}}\times\vect{r}+m\,\vect{\omega}\times(\vect{\omega}\times\vect{r})
+2m\,\vect{\omega}\times\vect{v}_{\rm rot}\,\,.
\end{equation}
Here, $\vect{\omega}=(\alpha-1)\dot{\varphi}_1\,\zhat$ is the rotation rate of the revolving axes,
$\vect{r}={\rho}\,\rhohat$ is the radial position of body 2, and
$\vect{v}_{\rm rot} = \vect{v} - (\vect{\omega}\times\vect{r})$ is the tangential velocity of body 2 in the revolving
frame, with $\vect{v}=\dot{\rho}\rhohat+\alpha\rho\dot{\varphi}_1\phihat$ 
being its tangential velocity in the static frame as defined by Eq.~\eqref{eq:velo}.
We note 
that $\vect{\omega}$ is not constant here, so the so-called ``Euler force'' term $m\dot{\vect{\omega}}\times\vect{r}$
cannot be excluded. Carrying out the cross-product algebra (see Appendix~\ref{appB}) yields
\begin{equation}
\label{eq:F2-1}
\vect{F}_2(\rho) = \vect{F}'_2(\rho) + m(\alpha-1)[\rho\ddot{\varphi}_1+2\dot{\rho}\dot{\varphi}_1]\,\phihat - 
m\rho\dot{\varphi}_1^2[(\alpha-1)^2+2(\alpha-1)]\,\rhohat\,\,.
\end{equation}
As before, we recognize the $\phihat$-term as being zero for any central force, and the $\rhohat$ term reduces to give
\begin{equation}
\label{eq:F2-2}
\vect{F}_2(\rho)-\vect{F}_1(\rho) = -\frac{\ell_1^2}{m\rho^3}(\alpha^2-1)\,\rhohat\,\,,
\end{equation}
where we have used $\dot{\varphi}_1=\ell_1/m\rho^2$ and have substituted $\vect{F}'_2(\rho) = \vect{F}_1(\rho)$
from Eq.~\eqref{eq:Forces}.
Thus, we have found the exact expression for the inverse
cube term defined in Eq.~\eqref{eq:invcube}. If $\alpha > 1$, this term is attractive and the orbit precesses
in the same direction as the body's orbital path; if $\alpha < 1$, this term is repulsive and the
orbit precesses in the opposite direction as the body's orbital path. For either case, 
the closer $\alpha$ is to unity, the less drastic the effect.

The physical interpretation of this extra force can be conceptually understood using uniform circular
motion as an example. If we consider two bodies of identical mass subjected to the same force, 
but moving at different rotational
speeds, then the two bodies must be at different radial distances from the force center. 
However, if we want to enforce the condition that the two bodies orbit at the same radial distance, 
then the force acting on the faster object must be increased. Specifically, the amount of force 
must be increased by the magnitude of the result given in Eq.~\eqref{eq:F2-2}.

We can demonstrate this generally if we first express the radial motion equation from Newton's second law as:
\begin{equation}
\label{eq:Newton2}
m\ddot{\rho}\,\rhohat = \vect{F}(\rho) + \frac{\ell^2}{m\rho^3}\,\rhohat\,\,,
\end{equation}
where the second term on the right hand side is often called the ``fictitious'' centrifugal force. We point out that 
any change to the angular momentum $\ell$ here will affect both the angular motion and the radial motion. 
That is, if we let 
$\ell\rightarrow\alpha\ell$, then $\dot{\varphi}\rightarrow\alpha\ell/m\rho^2$, but the centrifugal force term 
in Eq.~\eqref{eq:Newton2} also changes, therefore the solution for $\rho(t)$ will change. However, if we simultaneously
allow $\vect{F}(\rho)\rightarrow\vect{F}(\rho)-\ell^2(\alpha^2-1)/m\rho^3$ from Eq.~\eqref{eq:F2-2}, then we see that 
the effect on the radial motion due to the angular momentum change is canceled by the corresponding change in force:
\begin{align}
m\ddot{\rho}\,\rhohat &= \left(\vect{F}(\rho) 
- \frac{\ell^2}{m\rho^3}(\alpha^2 - 1)\,\rhohat\right) + \frac{\alpha^2\ell^2}{m\rho^3}\,\rhohat \nonumber\\
                      &= \vect{F}(\rho) + \frac{\ell^2}{m\rho^3}\,\rhohat\,\,.
\end{align}
Thus, the radial motion is preserved.
The angular motion, however, still remains changed: In the case of circular motion, the rotational speed will be different;
for other orbits, such as an ellipse, the difference in the angular motion manifests as precession.
\section{Finding the Precessing Orbit}
\label{sec:orbits}
To determine the orbit of the precessing body 2 as viewed from the static frame, we begin with Eq.~\eqref{eq:Binet},
and we let the total angular momentum of the precessing body be $\ell_2 = \alpha\ell_1$:
\begin{equation}
\label{eq:Binet-Rev0}
-\frac{(\alpha\ell_1)^2u^2}{m}\left(\frac{d^2u}{d\varphi^2} + u\right)\,\rhohat = \vect{F}_2(1/u)\,\rhohat\,\,.
\end{equation}
From Eq.~\eqref{eq:F2-2}, the total force acting on the precessing body 2 in the static frame
is given by the force that acts on the static orbit body 1 (assumed to be the inverse-square force) 
plus the inverse-cube force:
\begin{equation}
\label{eq:F2-3}
\vect{F}_2(1/u)\,\rhohat = -\beta u^2\,\rhohat-\frac{\ell_1^2}{m}(\alpha^2-1)u^3\,\rhohat\,\,,
\end{equation}
and thus
\begin{equation}
\label{eq:BinetRev}
\frac{(\alpha\ell_1)^2u^2}{m}\left(\frac{d^2u}{d\varphi^2} + u\right) = \beta u^2+\frac{\ell_1^2}{m}(\alpha^2-1)u^3\,\,.
\end{equation}
\begin{figure}
\includegraphics[width=6in]{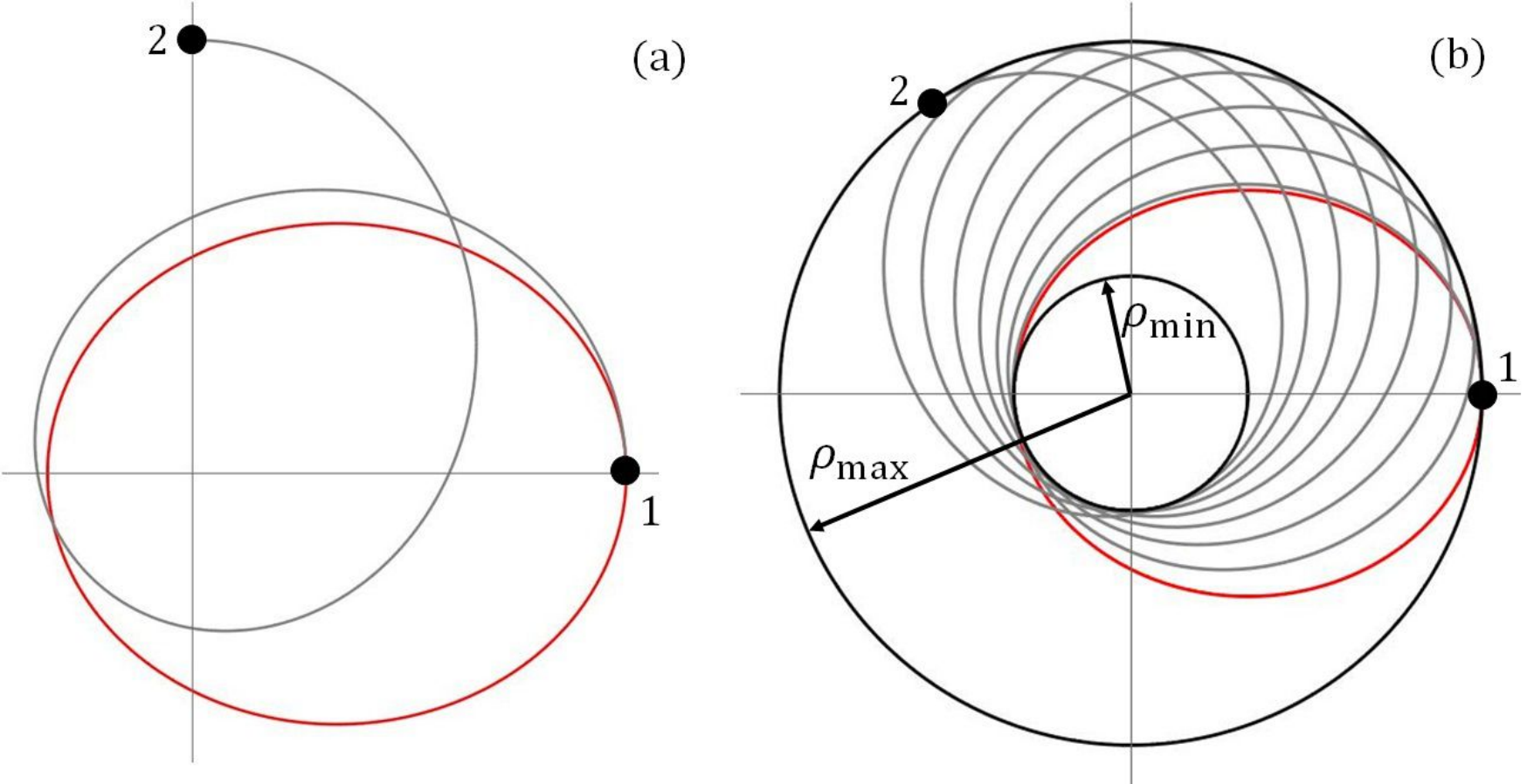}
\caption{(Color online) Plots of the static ellipse (red) and revolving ellipse (gray) for two different $\alpha$
         values. In panel (a), $\alpha = 5/4$, and we see that body 2 has revolved to $\varphi_2=\pi/2$ in the time
				 it takes body 1 to return to its starting position, but both bodies are at the same radial distance
				 $\rho=\rho_{\rm max}$. In panel (b), $\alpha = 21/20$ and we
				 see a less drastic precession for body 2. The two circles with radii $\rho=\rho_{\rm min}$ and 
				 $\rho=\rho_{\rm max}$ verify 
				 that the radial bounds for the two orbits are identical.				
				}
\label{fig:alphas}
\end{figure}
To solve Eq.~\eqref{eq:BinetRev}, we first reduce it to
\begin{equation}
\label{eq:BinetRev-2}
\frac{d^2u}{d\varphi^2} + u = \frac{m\beta}{(\alpha\ell_1)^2} + \frac{(\alpha^2-1)}{\alpha^2}u\,\,,
\end{equation}
which can be further reduced and rearranged to give
\begin{equation}
\label{eq:BinetRev-3}
\frac{d^2u}{d\varphi^2} + \frac{u}{\alpha^2} = \frac{m\beta}{(\alpha\ell_1)^2}\,\,.
\end{equation}
We note that this has the same form as Eq.~\eqref{eq:Binet-Kep}, and the solution is thus found to be 
\begin{equation}
\label{eq:revorb}
\rho_2(\varphi) =
\frac{\ell_1^2/(\beta\,m)}{1+\varepsilon\cos{\left(\dfrac{\varphi}{\alpha}\right)}}\,\,,
\end{equation}
where the eccentricity $\varepsilon$ is the same as Eq.~\eqref{eq:keporb}.
We note, then, that this expression is identical to Eq.~\eqref{eq:keporb}, except for the argument of the cosine, which 
accounts for the orbit's precession. In Fig.~\ref{fig:alphas}, we plot Eqs.~\eqref{eq:keporb} and~\eqref{eq:revorb}
for the case of an ellipse, with different $\alpha$ values to show its effect.
In panel (a) we have set $\alpha=5/4$ to exaggerate the 
precession and note that in one orbital period, body 1 has returned to its starting position but body 2 has
precessed to $\varphi_2=\pi/2$, as measured from the static frame. However, the radial distances at these two
positions are identical
($\rho=\rho_{\rm max}$) as expected. In panel (b), we have set $\alpha=21/20$ and show the results for seven orbital
periods, noting that the radial motion for the revolving ellipse is confined to the same
$\rho_{\rm min}\le\rho\le\rho_{\rm max}$ bounds as the static ellipse; this is verified by plotting 
circles centered at the focal point of the ellipse, with radii $\rho_{\rm min}$
and $\rho_{\rm max}$.

We note that all of the arguments laid out in this paper assume that body 1 is in a static orbit, 
while body 2 is in a precessing orbit. This is, of course, a consequence of our initial choice for the 
inertial reference frame. It is completely valid for one to consider body 2 as being in the 
static orbit, while body 1 precesses. In the Supplemental Materials, we consider 
this case in detail.~\cite{supplemental}
\section{Conclusion}
\label{sec:conc}
In this paper, we have presented Newton's \textit{Theorem of Revolving Orbits} using modern mathematical conventions,
and we have provided context for how it relates to standard topics in physics, such as the Kepler problem and 
rotating coordinate systems.
In our 
opinion this is a valuable extension of introductory orbital mechanics that can easily be incorporated into the
standard curriculum. For example, a common homework problem in various levels of mechanics 
courses~\cite{Goldstein,MT,Whittaker,Symon} is to pose
an inverse-cube ``perturbation term'' for the Kepler problem, i.e.
\begin{equation*}
F(r) = -\dfrac{\beta}{r^2}-\dfrac{\lambda}{r^3}\hspace{.5in}\beta,\lambda > 0\,\,,
\end{equation*}
and ask students to show that the solution for this modified force is a precessing ellipse, if certain conditions
are satisfied. Students are expected to use Eq.~\eqref{eq:Binet} to come up with a solution that 
``takes the form'' of a conic section that precesses at a $\lambda$-dependent rate.

However, when presented in the abstract like this there is very little \textit{physics} involved; 
it's effectively just a math problem for the students to solve. 
They do not gain any insight into \textit{why}
the inverse-cube term (and only an inverse-cube term) results in this particular precession. 
Given the work presented in this paper, we believe that there is an opportunity here to expose the students to an 
interesting, historically-important problem that they can solve using familiar concepts and techniques, thus allowing them to
build physical intuition. Such problems are not as abundant as we would like them to be, and as educators we
need to take advantage when they present themselves.
Once the precession argument is understood, connecting the $\lambda$ parameter to the body's angular momentum,
allowing $\lambda$ to vary, and observing the results (precession or spiral orbits) can then be 
explored in more detail with confidence.

Finally, we feel that using the rotating frame transformation to find the perceived force acting on the orbiting
body serves as an interesting application of such non-inertial reference frames.
While we have explicitly shown two such reference frame transformations as examples (see Supplemental Materials
for the second example~\cite{supplemental}), there are many others that
could be chosen. Expecting students to be able to find the transformed force and then the perceived orbit would 
be a reasonable exercise.
\appendix
\section{Recasting the Differential Equation}
\label{appA}
Newton's second law for a central force is given in polar coordinates by:
\begin{equation}
\label{eq:appNewton2}
m\left(\ddot{\rho}-\rho\,\dot{\varphi}^2\right)\,\rhohat = \vect{F}(\rho)\,\,.
\end{equation}
Often times, however, we are concerned with the shape of the orbit $\rho(\varphi)$ and not necessarily the time-dependence
of the polar coordinates $\rho(t)$, $\varphi(t)$.
This is easily achieved by making the substitution $\rho = 1/u$ and utilizing the chain rule:
\begin{equation}
\label{eq:appchainrule0}
\frac{d}{dt} = \frac{d}{d\varphi}\frac{d\varphi}{dt} 
= \frac{d}{d\varphi}\dot{\varphi} = \frac{\ell}{m\rho^2}\frac{d}{d\varphi}\,\,.
\end{equation}
Applying Eq.~\eqref{eq:appchainrule0} to $\rho$ gives
\begin{equation}
\label{eq:appchainrule}
\dfrac{d\rho}{dt}=\frac{\ell}{m\rho^2}\frac{d\rho}{d\varphi}\,\,.
\end{equation}
We then make the substitution $\rho = 1/u$ to the right hand side of Eq.~\eqref{eq:appchainrule}, which gives
\begin{equation}
\label{eq:apprhodot}
\dot{\rho} = \dfrac{\ell}{m}u^2\,\left(-\frac{1}{u^2}\frac{du}{d\varphi}\right) = -\frac{\ell}{m}\,\dfrac{du}{d\varphi}\,\,.
\end{equation}
We now apply Eq.~\eqref{eq:appchainrule0} to Eq.~\eqref{eq:apprhodot}, which gives
\begin{equation}
\label{eq:apprhodot2}
\dfrac{d\dot{\rho}}{dt} = -\dfrac{\ell}{m}\,\left(\dfrac{d}{d\varphi}\,\dfrac{d\varphi}{dt}\right)\,\dfrac{du}{d\varphi} 
=-\dfrac{\ell}{m}\,\left(\dfrac{\ell}{m\rho^2}\,\dfrac{d}{d\varphi}\right)\,\dfrac{du}{d\varphi}
=-\dfrac{\ell^2 u^2}{m^2}\,\dfrac{d^2u}{d\varphi^2}\,\,,
\end{equation}
where we again used $\rho = 1/u$ in the last expression.
Finally, inserting Eq:~\eqref{eq:apprhodot2} into Eq~\eqref{eq:appNewton2} with $\dot{\varphi}=\ell u^2/m$ leaves:
\begin{equation*}
-\dfrac{\ell^2u^2}{m}\left(\dfrac{d^2u}{d\varphi^2}+u\right)\,\rhohat = \vect{F}(1/u)\,\,,
\end{equation*}
which is Eq.~\eqref{eq:Binet}.
\section{The Rotating Frame Transformation}
\label{appB}
To define the force $\vect{F}_2(\rho)$ that acts on body 2 as viewed from the static frame, we transform into a frame that
rotates at the non-constant rate $\vect{\omega} = (\alpha-1)\dot{\varphi}_1\,\zhat$. In this rotating frame, the 
effective force on body 2 is the same force that acts on body 1 in the static frame, i.e. 
$\vect{F}'_2(\rho) = \vect{F}_1(\rho)$.
The relevant transformation is given by Eq.~\eqref{eq:transform}, with $\vect{r}=\rho\,\rhohat$,
$\vect{v}_{\rm rot}=\vect{v}-(\vect{\omega}\times\vect{r})$, and
$\vect{v}=\dot{\rho}\,\rhohat+\alpha\rho\dot{\varphi}_1\,\phihat$, as defined
in the main text. The relevant 
vector definitions and operations are thus given by:
\begin{align}
\vect{\omega}\times\vect{r} &= \rho(\alpha-1)\dot{\varphi}_1\,\phihat\\
\vect{v}_{\rm rot}=\vect{v}-(\vect{\omega}\times\vect{r}) &= \dot{\rho}\,\rhohat + \rho\dot{\varphi}_1\,\phihat\\
\label{eq:appcoriolis}
2m\,\vect{\omega}\times\vect{v}_{\rm rot} &= 
2m\dot{\rho}(\alpha-1)\dot{\varphi}_1\,\phihat - 2m\rho(\alpha-1)\dot{\varphi}_1^2\,\rhohat\\
\label{eq:appeuler}
m\,\vect{\dot{\omega}}\times\vect{r} &= m\rho(\alpha-1)\ddot{\varphi}_1\,\phihat\\
\label{eq:appcentrifugal}
m\,\vect{\omega}\times(\vect{\omega}\times\vect{r}) &= -m\rho(\alpha-1)^2\dot{\varphi}_1^2\,\rhohat
\end{align}
Using the above expressions in Eq.~\eqref{eq:transform} yields Eq.~\eqref{eq:F2-1}:
\begin{equation*}
\vect{F}_2(\rho) = \vect{F}'_2(\rho) + m(\alpha-1)[\rho\ddot{\varphi}_1+2\dot{\rho}\dot{\varphi}_1]\,\phihat - 
m\rho\dot{\varphi}_1^2[(\alpha-1)^2+2(\alpha-1)]\,\rhohat\,\,,
\end{equation*}
which we reduce in the main text to give Eq.~\eqref{eq:F2-2}:
\begin{equation*}
\vect{F}_2(\rho)-\vect{F}_1(\rho) = -\frac{\ell^2}{m\rho^3}(\alpha^2-1)\,\rhohat\,\,.
\end{equation*}

\begin{acknowledgments}
The authors thank the reviewers for their thoughtful comments and useful suggestions.
\end{acknowledgments}
The authors have no conflicts to disclose.
%

\end{document}